\documentclass[12pt,a4]{article}

\usepackage{amsthm,amsmath,latexsym,amssymb,amsfonts,amscd}
\usepackage{graphics,lscape,fancyhdr,array,stmaryrd,euscript}
\usepackage{empheq}
\usepackage{dsfont}
\usepackage{verbatim}
\usepackage{color,tikz,tikz-cd}
\usetikzlibrary[snakes]
\usepackage{relsize,slashed}
\numberwithin{equation}{section}
\usepackage{hyperref}
\newcommand{\pl}{\partial}
\newcommand{\pd}{\partial}

\usepackage{hyperref}

\usepackage[numbers,sort&compress]{natbib}
\usepackage[nottoc]{tocbibind}

\newcommand{\fud}[2]{{}^{#1}{}_{#2}\,}
\newcommand{\fdu}[2]{{}_{#1}{}^{#2}\,}

\newcommand{\besubeqs}{\begin{subequations}}
\newcommand{\esubeqs}{\end{subequations}}

\newcommand{\momega}{{\boldsymbol{\omega}}}
\newcommand{\mC}{{\boldsymbol{C}}}

\newcommand{\hs}{{\ensuremath{\mathfrak{hs}}}}
\newcommand{\hsdouble}{{\ensuremath{{}^\pi\mathfrak{hs}}}}

\begin{document}
\pagenumbering{gobble}
\hfill
\begin{flushright}
    {}
\end{flushright}
\vskip 0.01\textheight
\begin{center}
{\Large\bfseries 
Towards Massless Sector of Tensionless Strings on $AdS_5$
\vspace{0.4cm}}

\vskip 0.03\textheight

Alexey \textsc{Sharapov}${}^{1}$, Evgeny \textsc{Skvortsov}${}^{2,3}$ and Tung \textsc{Tran}${}^{2,4}$

\vskip 0.03\textheight

{\em ${}^{1}$Physics Faculty, Tomsk State University, \\Lenin ave. 36, Tomsk 634050, Russia}\\
\vspace*{5pt}
{\em ${}^{2}$ Albert Einstein Institute, \\
Am M\"{u}hlenberg 1, D-14476, Potsdam-Golm, Germany}\\
\vspace*{5pt}
{\em ${}^{3}$ Lebedev Institute of Physics, \\
Leninsky ave. 53, 119991 Moscow, Russia}\\
\vspace*{5pt}
{\em ${}^{4}$ Arnold Sommerfeld Center for Theoretical Physics\\
Ludwig-Maximilians University Munich\\
Theresienstr. 37, D-80333 Munich, Germany}

\end{center}

\vskip 0.02\textheight

\begin{abstract}
{A} Higher Spin Gravity in five dimensions is constructed. It was shown recently that constructing formally consistent classical equations of motion of higher spin gravities is equivalent to finding a certain deformation of a given higher spin algebra. A strong homotopy algebra encoding the interaction vertices then follows. We propose two different and novel realizations of the deformed higher spin algebra in the case of five dimensions: one in terms of the universal enveloping algebra of $su(2,2)$ and the other by means of oscillator variables. Both the new realizations admit supersymmetric extensions and the $\mathcal{N}=8$ case underlies the massless sector of tensionless strings. 
\end{abstract}
\newpage
\tableofcontents
\newpage
\section{Introduction}
\pagenumbering{arabic}
\setcounter{page}{2}
The Higher Spin Gravity is one of the approaches to the problem of quantum gravity whose main idea is to extend the metric field with (usually infinitely many) massless fields of spin greater than two. A rich gauge symmetry associated with the higher spin extension is expected to render higher spin gravities renormalizable and even finite, thereby serving as an alternative and/or companion of supersymmetry. The conjectural `no counterterm' argument downgrades the problem of quantum consistency to a much simpler task of constructing a purely classical higher spin gravity. The crux, however, is a zoo of no-go theorems that makes it hard to find such theories in the landscape of conventional field theories. Many of these theorems are about the flat space background \cite{Weinberg:1964ew,Coleman:1967ad,Bekaert:2010hp}. Recently, the most important of these no-go's have found their anti-de Sitter space cousins.\footnote{For example, the Weinberg and Coleman--Mandula theorems imply, basically, that $S=1$ if massless higher spin particles are present as asymptotic states. Likewise, asymptotic higher spin symmetry at the boundary of $AdS$ trivializes the holographic $S$-matrix \cite{Maldacena:2011jn,Boulanger:2013zza,Alba:2013yda,Alba:2015upa}, which has to be given by a free CFT's correlators (AdS/CFT analog of $S=1$ in flat space). Within the local field theory approach, the flat space non-locality \cite{Bekaert:2010hp} has found its $AdS$ counterpart \cite{Bekaert:2015tva,Sleight:2017pcz,Ponomarev:2017qab}.} 

At present there are only three examples of higher spin gravities that avoid all the no-go theorems one way or another: (i) extension of the Chern--Simons formulation of $3d$ gravity \cite{Blencowe:1988gj,Campoleoni:2010zq,Henneaux:2010xg}, which has been a very useful toy-model over the last years \cite{Gaberdiel:2012uj}; (ii) conformal higher spin gravity \cite{Segal:2002gd,Tseytlin:2002gz,Bekaert:2010ky} where the non-locality is tamed by the local Weyl symmetry and it shares many features with conventional lower spin theories; (iii) chiral higher spin gravity, which exists both in $4d$ flat \cite{Metsaev:1991nb,Metsaev:1991mt,Ponomarev:2016lrm,Ponomarev:2017nrr,Skvortsov:2018jea} and anti-de Sitter \cite{Skvortsov:2018uru} spaces. As shown in \cite{Skvortsov:2018jea}, the chiral higher spin gravity, being consistent at the quantum level, has $S=1$ in flat space and a non-trivial $S$-matrix in $AdS_4$. The latter can be used for deriving correlation functions in Chern--Simons Matter theories \cite{Skvortsov:2018uru}.

The AdS/CFT correspondence gives an important handle on the higher spin theories since the simplest CFT duals thereof are just free CFT's \cite{Sundborg:2000wp,Sezgin:2002rt,Klebanov:2002ja}. Free (or weakly coupled) CFT's do not have a large gap in the dimensions of single-trace operators and hence the existence of the gravitational dual is debatable \cite{Heemskerk:2009pn}. The main difficulty is that the simplest holographic higher spin models cannot be conventional field theories due to severe nonlocalities required by the higher spin symmetry \cite{Bekaert:2015tva,Sleight:2017pcz,Ponomarev:2017qab}. Yet, the existence of CFT dual descriptions allows one to address some problems of higher spin gravities from the boundary vantage point.

The free limit of $\mathcal{N}=4$ SYM is supposed to be dual to the tensionless limit of the IIB string theory on $AdS_5\times S^5$ \cite{HaggiMani:2000ru,Sundborg:2000wp}. A priori there are no reasons to expect the tensionless limit be well-defined. Indeed, it is quite singular in the flat space background \cite{Gross:1988ue}. When the background has the $AdS_{d+1}$ factor the tensionless limit corresponds to very long strings $l_s \gg R$, the length $l_s$ being very large compared to the AdS radius $R$, see e.g. \cite{Tseytlin:2002ny, Bonelli:2003zu} for discussion. The Planck length is still assumed to be small, so that we are dealing with the weakly coupled string theory. The absence of any argument for why (or under which further conditions) the tensionless limit has to admit a simple, possibly, weakly-coupled description is another side of the same problem. In this respect it comes as a surprise that there is a worldsheet description of the tensionless strings on $AdS_3$ \cite{Eberhardt:2018ouy} and there is some understanding of the holographic fishnet models \cite{Gromov:2019aku}.\footnote{Fishnet theories admit a weakly-coupled limit where the higher spin currents have small anomalous dimensions. Therefore, fishnet theories are similar to weakly-coupled SYM. The higher spin currents in these CFT's are dual to massless higher spin fields in $AdS_5$. }

Bearing the holographic higher spin theories in mind, it is natural to look for structures that survive in  presence of nonlocalities.  An infinite-dimensional extension of conformal symmetry $so(d,2)$, known as a higher spin symmetry, is one of such structures.  The higher symmetries correspond to higher rank conserved tensors present in any free CFT (including the free SYM). Every free CFT comes equipped with a Higher Spin Algebra --- the symmetry algebra of the free equations of motion \cite{Eastwood:2002su}. As such, higher spin symmetries always arise from associative algebras\footnote{On can multiply any two symmetries of a linear equation, but, in general, only the commutator of the two is a symmetry for nonlinear equations.} and are closely related to the universal enveloping algebra of the conformal algebra $so(d,2)$ (and its supersymmetric extensions).

Being identified with a global symmetry on the CFT side, the higher spin algebra carries complete information about the spectrum of single-trace operators and their correlators.\footnote{Indeed, the correlation functions are just the simplest higher spin algebra invariants \cite{Colombo:2012jx,Didenko:2013bj,Didenko:2012tv,Bonezzi:2017vha}. } It has to be gauged in the gravitational dual producing thus inevitable nonlocalities. There is, however, an approach that makes the higher spin problem well-defined mathematically, but with some sacrifice in other features of conventional field theory. It focuses upon construction of formally consistent classical equations of motion. The equations of formal higher spin gravities read
\begin{align}\label{mastereq}
    d\Phi&=\mathcal{V}_2(\Phi,\Phi)+\mathcal{V}_3(\Phi,\Phi,\Phi)+\cdots \,, &&dd\equiv0\,,
\end{align}
where $\Phi$ is a master-field to be defined later. The equations are very close in spirit to those of String Field Theory, see e.g. \cite{Gaberdiel:1997ia,Erler:2013xta}. It is also true that strong homotopy algebras, more precisely certain $L_\infty$- and $A_\infty$-algebras  \cite{Gaberdiel:1997ia,Erler:2013xta,Sharapov:2018kjz,Sharapov:2019vyd}, play a crucial role in both the theories. For the higher spin case, the bilinear vertex $\mathcal{V}_2$ is completely determined $\mathcal{V}_2=\Phi\star \Phi$ by the product $\star$ in a given higher spin algebra, so that the problem is to construct all higher vertices $\mathcal{V}_n$ in a consistent way. There is a handful of such models known \cite{Vasiliev:1990en,Prokushkin:1998bq,Vasiliev:2003ev,Sagnotti:2005ns,Bonezzi:2016ttk,Bekaert:2017bpy,Arias:2017bvi,Grigoriev:2018wrx,Sharapov:2019vyd} that take advantage of various techniques and approaches to the problem. We do this for the $5d$ case.

The study of five-dimensional higher spin gravities is well motivated by the relation to string theory. There should be a well-defined massless subsector of the tensionless\footnote{Note that \cite{Sezgin:2002rt} discusses a certain critical value of the tension.} string theory (dual to the bilinear single-trace operators of the free SYM) that is described by $\mathcal{N}=8$ higher spin gravity. The important steps towards this theory were made in \cite{Sezgin:2001yf}, where the higher spin superalgebra and the free equations were given. Options with less supersymmetries are also possible, including the purely bosonic theory, also called Type-A for any $d$ \cite{Vasiliev:2003ev,Sagnotti:2005ns,Bekaert:2017bpy,Arias:2017bvi}. The Type-A theory should be regarded as an important building block of supersymmetric models. For example, the known supersymmetric models in $AdS_4$ \cite{Sezgin:2012ag} can be thought of as supersymmetric extensions of the bosonic theory. Likewise, at least kinematically, there is a family of higher spin extensions of $(p)su(2,2|N)$ for any $N$ that is relevant for the theories in $AdS_5$ \cite{Fradkin:1989yd,Sezgin:2001yf,Fernando:2009fq,Govil:2013uta}. As different from \cite{Vasiliev:2003ev,Sagnotti:2005ns,Bekaert:2017bpy,Arias:2017bvi} the realizations of the $5d$ higher spin algebra we use in the paper admit a simple supersymmetric extension.

Constructing formal Higher Spin Gravity in five dimensions has been an open problem since the late 1990s. The relevant higher spin algebra had been known \cite{Fradkin:1989yd}. Free fields, including the mixed-symmetry ones, were studied in \cite{Metsaev:2002vr,Metsaev:2004ee,Metsaev:2014sfa}. Certain cubic vertices for the $\mathcal{N}=0,1,2$ cases were constructed in \cite{Vasiliev:2001wa,Alkalaev:2002rq,Alkalaev:2010af}. The free equations of type \eqref{mastereq} were analyzed in \cite{Sezgin:2001zs}. In Sec. \ref{sec:deformation}, we explain why all the previously known methods do not work here when it comes to interactions. Our solution heavily relies on the work \cite{Sharapov:2019vyd} that reduces the problem of constructing the interaction vertices to a much simpler problem of deforming a certain extension of a given higher spin algebra. We found two different and novel ways to construct such a deformation, which should have a wider range of applications. 

The first one appeals to the very definition of higher spin algebras as quotients of universal enveloping algebras. The ideal to be quotiented out is the annihilator of the corresponding free field and is also known as the Joseph ideal. The ideal is generated by a few quadratic relations. It turns out that one can simply deform these relations together with the commutator of the translation generators. This leads to an interesting way to deform (quotients of) universal enveloping algebras. The second one takes advantage of the quasi-conformal realizations \cite{Gunaydin:2000xr,Fernando:2009fq,Govil:2013uta} that were previously underrated in the higher spin context. The main feature is that they resolve all of the Joseph relations and give the minimal oscillator realization of the free field and of the corresponding higher spin algebra. We found a way to deform the quasi-conformal realization so that the deformed Joseph's relations are satisfied. 

The outline of the paper is as follows. In Section \ref{sec:hsalg}, we review various definitions of the $5d$ higher spin algebra and discuss the free equations, i.e., up to $\mathcal{V}_2$ in \eqref{mastereq}. In Section \ref{sec:vertices}, we summarize the central result of \cite{Sharapov:2019vyd} that simplifies the problem. The main results of the present paper are in Section \ref{sec:deformation}, where we formulate two different ways to describe the deformed higher spin algebra. In Section \ref{sec:Einstein}, we show that the peculiar deformation of the conformal algebra's commutators gives rise to the Einstein equations and is therefore well-motivated. The discussion of the obtained results can be found in Section \ref{sec:conclusions}.

\section{\label{sec:hsalg}Initial Data}
The only input is given by a Higher Spin Algebra. The Type-A theory is, by definition, the anti-holographic dual to the large-$N$ free (or critical if $d\neq 4$) vector model \cite{Klebanov:2002ja}. For the $AdS_5/CFT_4$ case the higher spin algebra, we shall call $\hs$, comes from the universal enveloping algebra of $su(2,2)$, whose  generators $T\fdu{A}{B}$, $A,B=1,\ldots,4$ obey\footnote{The indices $A,B,\dots$ are the indices of the (anti)-fundamental representation of $su(2,2)$.}
\begin{align}\label{T}
    [T\fdu{A}{B},T\fdu{C}{D}]&= \delta_A{}^D\, T\fdu{C}{B}-\delta_C{}^B\, T\fdu{A}{D}\,.
\end{align}
The higher spin algebra is defined then to be the quotient of the universal enveloping algebra $U(su(2,2))$ by a certain two-sided ideal \cite{Eastwood:2002su}, known as the Joseph ideal $J$, $\hs=U(su(2,2))/J$. The state space $|\phi\rangle$ of a free $4d$ conformal scalar field, $\square \phi=0$, is the minimal unitary irreducible representation of $su(2,2)$ and the Joseph ideal is the annihilator of this module. The ideal is generated by the quadratic relations\footnote{\label{ft:trace}Here, $[AB]$ means that one has to add minus the same term with $AB$ swapped. We also used the first two to simplify the third one, otherwise a projector onto $(0,2,0)$ has to written down explicitly.}
\besubeqs\label{Joseph}
\begin{align}\label{JosephA}
    C_2&=T\fdu{A}{B}T\fdu{B}{A}=-3\,,\\
    T\fdu{A}{C}T\fdu{C}{B}&=-2\, T\fdu{A}{B}+\frac14 C_2\, \delta\fdu{A}{B}\,,\label{JosephB}\\
    \{T\fdu{[A}{[B}, T\fdu{C]}{D]}\}&=\delta\fdu{A}{B}\delta\fdu{C}{D}-\delta\fdu{C}{B}\delta\fdu{A}{D}\label{JosephC}\,.
\end{align}
\esubeqs
Elements of the higher spin algebra $\hs$ are polynomials $f(T)$ (or formal power series) in $T\fdu{A}{B}$ modulo the Joseph relations. It is easy to see that the Joseph relations wash away all $su(2,2)$-tensors except for
\begin{align}\label{hselem}
    f&= \sum_k f_{B_1\ldots B_k}^{A_1\ldots A_k} \, T\fdu{A_1}{B_1}\cdots \, T\fdu{A_k}{B_k}\,,
\end{align}
where the coefficients are traceless and symmetric in upper and lower indices, i.e., define an irreducible representation of weight $(k,0,k)$.

From the conformal point of view, these representations are in one-to-one with conformal Killing tensors (in particular the $(1,0,1)$ is the adjoint of $su(2,2)$ associated with conformal Killing vectors). It can be shown \cite{Eastwood:2002su} that each homogeneous element of $\hs$ defines a conformal Killing tensor\footnote{Conformal Killing tensors obey $\pd^{a_1} v^{a_2\cdots a_{k+1}}+\text{permutations}-\text{traces}=0$. } $v^{a_1\cdots a_k}(x)$ and
\begin{align}
    \delta_v \phi(x)&= v^{a_1\cdots a_k} \pd_{a_1}\cdots \pd_{a_k} \phi(x)+\cdots 
\end{align}
is a (higher derivative) symmetry of $\square \phi(x)=0$. The equation being linear, the symmetries form an associative algebra, the one we have just described, see also \cite{Mikhailov:2002bp}.

One can take Rels. (\ref{T}) and (\ref{Joseph}) as an {\it ab initio} definition of the higher spin algebra $\hs$. In practical applications it is sometimes convenient to resolve (some of) the Joseph relations by passing to an appropriate realization. One common way to resolve some of the Joseph relations is to introduce two quartets of oscillator variables  $a_A$ and $b^B$ in the fundamental and anti-fundamental representations of $su(2,2)$ (they generate the Weyl algebra $A_4$):
\begin{align}
[a^A,a^B]=0\,,\quad [b_A,b_B]=0\,,\quad [a_A,b^B]= \delta\fdu{A}{B}\,.
\end{align}
Then the $su(2,2)$ generators are given by
\begin{align}\label{usualosc}
    T\fdu{A}{B}&=\frac12\{a_A,b^B\}-\frac14 \delta\fdu{A}{B} N\,, 
\end{align}
where the $u(1)$ generator $N=\frac12\{a_C,b^C\}$ commutes with $T\fdu{A}{B}$. The same $\hs$ can now  be defined as a subquotient of the oscillator algebra: 
\begin{align}\label{subquotient}
    \hs \ni f \;& &&\Leftrightarrow&& [f,N]_\star =0\,, && f\sim f+g\star N\,,
\end{align}
The first relation forces the monomials of $f(a,b)$ to have an equal number of $a$'s and $b$'s. The quotient with respect to $N$ makes the Taylor coefficients effectively traceless, as in \eqref{hselem}. It is this realization that was used in \cite{Sezgin:2001zs} to study the spectrum and free higher spin equations.

While  Rel. \eqref{JosephC} is identically satisfied (up to the traces, see footnote \ref{ft:trace}), the remaining two Joseph's relations are not and gauging of $u(1)$ in \eqref{subquotient} lands us onto the right algebra. There is also a way to resolve all Joseph's relations. It is known as a quasi-conformal realization \cite{Gunaydin:2000xr,Fernando:2009fq,Govil:2013uta}. The idea is to represent the higher spin algebra by a minimal possible number of oscillators.\footnote{In some sense, this is exactly the problem that Joseph addressed in \cite{Joseph:1974hr}: how to realize irreducible representations by the minimal number of oscillators.} In the case under consideration  the minimal number of canonical pairs is three.\footnote{Apart from the rigorous results by Joseph \cite{Joseph:1974hr} this can be seen by noticing that the solution space of the Klein-Gordon equation has functional dimension three. } Therefore, we set
\begin{align}
    [z,p_z]=[y,p_y]=[x,p_x]=i\,.
\end{align}
The most nontrivial statement of the quasi-conformal realization is that the following two composite operators\footnote{We note that the realization we present here is slightly different from the one in \cite{Fernando:2009fq,Govil:2013uta}.}
\begin{align}
    Y_L^A&=\Big\{z, p_z, 0, \frac{1}{x}(z p_y-p_z y-\tfrac12) \Big\}\,,&
    Y^A_R&=\big\{y,p_y,x,p_x\big\}
\end{align}
can be used to define the generators of $\hs$. Let us also define $Y_\pm^A=Y_L^A\mp iY_R^A$.
Then one can easily check that the operators 
\begin{align}\label{sufromY}
    T\fdu{A}{B}&= -\frac{i}{2}\left(Y_A^+ Y^B_- -\tfrac14 \delta\fdu{A}{B} Y_C^+ Y^C_-\right)
\end{align}
obey the commutation relations (\ref{T}) as well as  Joseph's relations \eqref{Joseph}. We note in passing that the most singular component of the $su(2,2)$-generators coincides with the Calogero Hamiltonian:
\begin{align}
    iL_{44}=H&= p^2 +\frac{Q}{x^2}\,,
\end{align}
where $Q$ is a quartic polynomial in the other variables. 
The rest of the $su(2,2)$ generators are either non-singular or involve $1/x$.\footnote{See \cite{Fernando:2009fq,Govil:2013uta} for the wave functions and complete construction of the unitary representation.}

An important role is played in what follows by the Lorentz subalgebra $so(4,1)\sim sp(4)$ ---  it is the maximal symmetry that remains undeformed. It allows one to split the $su(2,2)$-generators into the Lorentz generators $L_{AB}$ and translations $P_{AB}$:
\begin{align}\label{LorTrans}
    L_{AB}&= T_{A|B}+T_{B|A}\,, &
    P_{AB}&= T_{A|B}-T_{B|A}\,.
\end{align}
Hereafter all $sp(4)$-indices are raised and lowered with the help of $sp(4)$-invariant tensor $C_{AB}=-C_{BA}$. The $su(2,2)$ commutation relations \eqref{T} then read
\besubeqs\label{LLLPPP}
\begin{align}\label{LL}
    [L^{AB},L^{CD}]&= L^{AD}C^{BC}+L^{BD}C^{AC}+L^{AC}C^{BD}+L^{BC}C^{AD}\,,\\
    [L^{AB},P^{CD}]&= P^{AD}C^{BC}+P^{BD}C^{AC}-P^{AC}C^{BD}-P^{BC}C^{AD}\label{LP}\,,\\
    [P^{AB},P^{CD}]&= L^{AD}C^{BC}-L^{BD}C^{AC}-L^{AC}C^{BD}+L^{BC}C^{AD}\label{PP}\,.
\end{align}
\esubeqs
In order to write the free equations of motion we need the automorphism $\pi$ that any higher spin algebra is equipped with. It acts by altering the sign of translations, while leaving the Lorentz generators intact: $(\pi f)(T)=f(L,-P)$.

The master field $\Phi$ of the simplest higher spin gravities consists of 
two components:\footnote{This was first proposed in \cite{Vasiliev:1988sa} for the $4d$ system.} one-form $\omega$ and zero-form $C$, both taking values in a given higher spin algebra $\hs$. Loosely speaking, for $|\phi\rangle$ being the irreducible representation corresponding to the free scalar field $\phi(x)$, the higher spin-algebra $\hs$ is the space of all linear maps $|\phi\rangle\rightarrow |\phi\rangle$, i.e., $|\phi\rangle\langle \phi|$. The space of single-trace operators is $|\phi\rangle\otimes |\phi\rangle$. The latter is formally isomorphic to $\hs$ up to the conjugation that maps $|\phi\rangle$ to $\langle \phi|$. We want to gauge $\hs$ in the bulk since it is a global symmetry on the CFT side. The single-trace operators are dual to bulk physical states that are in $C$. We see that $\hs$ is formally isomorphic to the algebra of single-trace operators, which allows one to identify $C$ up to the $\pi$-map with $\hs$ as well. 

Given one or another realization of the higher spin algebra, the free equations of motion read
\begin{align}\label{freeeq}
    d\omega&= \omega\star \omega\,,&
    dC&= \omega \star C-C\star \pi(\omega)\,,
\end{align}
where $\star$ is the product in $\hs$ and $d$ is the exterior differential  $d=dx^\mu \pd_\mu$. 
The first equation describes the maximally symmetric higher spin backgrounds. The simplest one is $AdS_5$ and has only the spin-two subsector activated:
\begin{align}
    \omega&=\frac12 h^{AB}\, P_{AB}+ \frac12 \varpi^{AB}\, L_{AB}\,.
\end{align}
Here $h^{AB}=-h^{BA}$ is the f{\"u}nf-bein and $\varpi^{AB}=\varpi^{BA}$ is the spin-connection. The second equation in (\ref{freeeq}) describes the physical states that are dual to the single-trace operators of the free scalar CFT. The automorphism $\pi$ is responsible for  the conjugation   $|\phi\rangle\mapsto \langle \phi|$. These are the correct free equations that are completely fixed by representation theory. For the $AdS_5$ case they were studied in \cite{Sezgin:2001zs}. Without going into the details we just state that the spectrum of the theory is given by the massless fields on $AdS_5$ with all integer spins $s=0,1,2,3,\ldots$ \cite{Sezgin:2001zs}. Truncation to even spins and Yang-Mills gaugings with $u(N)$, $o(N)$, $usp(N)$ are also possible.

\section{\label{sec:vertices}Vertices}
The problem of formal higher spin gravities is to find a nonlinear
completion of the free system \eqref{freeeq}. Its general form reads
\besubeqs\label{problemgen}
\begin{align}\label{omegaeq}
d\omega&=\omega\star \omega + \mathcal{V}_3(\omega,\omega,C)+O(C^2)\,,\\
d C&=\omega\star C-C\star {\pi}(\omega)+\mathcal{V}_3(\omega,C,C)+O(C^3)\label{Ceq} \,,
\end{align}
\esubeqs
where the bilinear terms $\mathcal{V}_2$ are displayed explicitly. The interactions, $\mathcal{V}_n$, are constrained by the formal consistency that stems from the nilpotency of the exterior differential, $dd\equiv 0$.\footnote{One can think of the right-hand side as of the definition of an abstract differential $d$. This trick allows one to avoid the problem of nonlocality. Whenever $d$ is taken to be the exterior differential $d=dx^\mu \pl_\mu$ the equations become a set of PDE's that faces the nonlocality problem. } Intuitively, it is clear that the higher spin algebra $\hs$ (together with $\pi$) is the only input data for the problem and the interaction vertices $\mathcal{V}_n$, $n>2$, should be derivable from it. The precise relation between $\hs$ and $\mathcal{V}_n$'s for any $\hs$ was established in \cite{Sharapov:2018kjz,Sharapov:2019vyd}.

To formulate the result, let us extend the higher spin algebra with the $\pi$-automorphism by defining the smash-product algebra  $\hsdouble=\hs\rtimes \mathbb{Z}_2$, where $\mathbb{Z}_2=(1,k)$, $k^2=1$. The general element of $\hsdouble$ has the form  $a=a'\cdot 1+a''\cdot k$ for  $a',a''\in \hs$ and multiplication is given by 
\begin{align}
a\star b&=(a'b'+a''\pi(b''))\cdot 1 +(a''\pi(b')+a'b'')\cdot k\,,
\end{align}
i.e. $kxk=\pi(x)$, $x\in \hs$. Clearly, we can get the free equations \eqref{freeeq} by omitting $\pi$ in the second equation while requiring $\omega=\omega'\cdot 1$ and $C=C''\cdot k$. The same trick would work for the nonlinear equations \eqref{problemgen} as well. The smashed product allows us to incorporate the $\pi$-automorphism through the generator $k$. 

The main statement of \cite{Sharapov:2019vyd} is that (up to formal field re-definitions) the nontrivial interaction  vertices  are in one-to-one correspondence with the nontrivial  deformations of $\hsdouble$ as an associative algebra. The deformation of an associative algebra $\hsdouble$  appears to be a much simpler problem to solve.\footnote{In more technical terms \cite{Sharapov:2018kjz}, one can construct a certain strong homotopy algebra, $A_\infty$, from the one-parameter family of associative algebras. Higher spin algebras are just particular examples for which such a deformation is possible after extending them with the $\pi$-automorphism. } Let us assume for a moment that the deformation of $\hsdouble$ is known and is given by the formal series in the formal parameter we denote $\nu$
\begin{align}\label{circ}
    a\circ b&= a\star b +\phi_1(a,b)\nu +\phi_2(a,b)\nu^2 +\cdots\,,
\end{align}
so that $a\circ(b\circ c)=(a\circ b)\circ c$. Then, there are general formulas expressing all the vertices in a certain minimal form:\footnote{The form is minimal in the sense of having the least number of nontrivial structure maps that are, moreover, expressed entirely in terms of the deformed product. However, this form is less convenient to discuss reality conditions and various truncations, extensions.}
\besubeqs
\begin{align}
    \mathcal{V}_n(\omega,\omega,C,\ldots,C)&=+f_n(\omega,\omega,C,\ldots\;)\star C\,,\\
    \mathcal{V}_n(\omega,C,\ldots,C,C)&=+f_n(\omega,C,\ldots,C)\star C\,,\\
    \mathcal{V}_n(C,\omega,\ldots,C,C)&=-f_n(C,\omega,\ldots,C)\star C\,.
\end{align}
\esubeqs
The structure functions $f_n$ are given by the sum over graphs
\begin{align}
f_n(a,b,u,\ldots,w)&=\sum_\Gamma
\parbox{4cm}{\begin{tikzpicture}[scale=0.4]
    \def\R{0.20}
    \draw[black,thick] node[below]{$\scriptstyle a$} (0,0) -- (3.5,3.5);
    \draw[black,thick] (1,1) -- (2,0) node[below]{$\scriptstyle b$};
    \filldraw[color=black,fill=black]   (1,1)  circle (\R) node[left,xshift=-0.2cm]{$\scriptstyle m_k+1$};
    \draw[black,thick] (1.5,1.5) -- (2,1.0) node[below right]{$\scriptstyle u$};
    \draw[black,thick] (2,2) -- (2.5,1.5);
    \draw[black,thick] (2.5,2.5) -- (3,2);
    \draw[black,thick] (3,3) -- (4,2);
    \filldraw[color=black,fill=black]   (3,3)  circle (\R); 
    \draw[black,thick] (3.9,3.9) -- (6.5,6.5);
    \draw[black,thick] (4.5,4.5) -- (5,4);    
    \draw[black,thick] (5,5) -- (6,4);
    \filldraw[color=black,fill=black]   (5,5)  circle (\R) node[left,xshift=-0.2cm]{$\scriptstyle m_{1}+1$};    
    \draw[black,thick] (5.5,5.5) -- (6,5);
    \draw[black,thick] (6,6) -- (6.5,5.5);    
    \draw[black,thick] (6.5,6.5) -- (7,6) node[below right]{$\scriptstyle w$};
    \draw[snake=brace,raise snake=5pt,black, thick] (5.5,5.5) -- (6.5,6.5) node[left, xshift=-0.25cm, yshift=0.15cm]{$\scriptstyle l_1$};
    \draw[snake=brace,raise snake=5pt,black, thick] (1.5,1.5) -- (2.5,2.5) node[left, xshift=-0.25cm, yshift=0.15cm]{$\scriptstyle l_k$};   
\end{tikzpicture}}
\end{align}
where the edges represent the arguments, simple vertices represent the direct products (formally $\phi_0(a,b)=a\star b$) and labelled bold vertices represent $\phi_{m_i+1}(\bullet,\bullet)$. The sum is over all graphs such that the labels $(l_1,\ldots,l_k)$ and $(m_1,\ldots,m_k)$ form two nested Young diagrams with the first rows of equal length:
\begin{align}
    \begin{tikzpicture}[scale=0.4]
        \draw[thick, black] (0,0) -- (3,0) -- (3,1) -- (6,1) -- (6,2) -- (9,2) -- (9,2.5);
        \draw[thick, black] (0,0) -- (0,2.5);
        \draw[thick, black] (0,3.5) -- (0,6) -- (15,6) -- (15,5) -- (13,5) -- (13,4) -- (11,4) -- (11,3.5);
        \draw[black] (0,2.5) to [out=20,in=185] (9,2.5);
        \draw[black] (0,3.5) to [out=20,in=185] (11,3.5);
        \node[below] at (1.5,0) {$l_1$};
        \node[above] at (7.5,6) {$n-2-k$};
        \node[below] at (4.5,1) {$l_2$};
        \node[below] at (7.5,2) {$l_3$};
        \node[below] at (14,5) {$l_k$};
        \draw[<->] (-1,0) -- (-1,6);
        \node[left] at (-1,3) {$k$};
        \draw[fill=lightgray] (0.1,2.4) to [out=20,in=180] (8,2.35) -- (8,2.1) -- (5,2.1) -- (5,1.1) -- (2, 1.1) -- (2,0.1) -- (0.1, 0.1) -- (0.1,2.4) ;
        \node[above, yshift=-0.1cm] at (1,0.1) {$m_1$};
        \node[above, yshift=-0.1cm] at (3,1.1) {$m_2$};
        \draw[fill=lightgray]  (0.1,3.7) to [out=20,in=180] (10,3.6) -- (10,4.1) -- (12, 4.1) -- (12,5.1) -- (14.9, 5.1) -- (14.9, 5.9) -- (0.1, 5.9) ;
        \node[above, yshift=-0.1cm] at (14,5.1) {$m_k$};
        \node[above, yshift=-0.1cm] at (11,4.1) {$m_{k-1}$};
\end{tikzpicture}
\end{align}
For instance, with these prescriptions, one can find the following explicit expressions for the cubic, quartic, and quintic vertices:\footnote{For the $AdS_4$ case the first few vertices are very close to those discussed e.g. in \cite{Vasiliev:1988sa,Sezgin:2002ru}. }  
\begin{align}\label{cubic}
    \mathcal{V}_3(\omega,\omega,C)&=\phi_1(\omega,\omega)\star C\,,\\[2mm]
    \mathcal{V}_4(\omega,\omega,C,C)&=\phi_2(\omega,\omega)\star C\star C +\phi_1(\phi_1(\omega,\omega),C)\star C\,,\notag \\[2mm]
    \mathcal{V}_5(\omega,\omega,C,C,C)&=\phi_3(\omega,\omega)\star C\star C\star C+\phi_2(\phi_1(\omega,\omega),C)\star C\star \notag C+\phi_1(\phi_2(\omega,\omega),C)\star C\star C\\&+\phi_1(\phi_2(\omega,\omega)\star C,C)\star C+\phi_1(\phi_1(\phi_1(\omega,\omega),C),C)\star C\,.\notag
\end{align}
There is also a simple differential equation that allows one to generate the vertices $\mathcal{V}_n$ and to prove their formal consistency \cite{Sharapov:2018kjz,Sharapov:2019vyd,Sharapov:2019qqq}. Thus, we see that the vertices are completely determined by the associative $\circ$-product (\ref{circ}).

It was also shown in \cite{Sharapov:2019vyd} that the equations of the formal Higher Spin Gravities are integrable. Solving the (highly nonlinear) equations \eqref{problemgen} is equivalent to solving a much simpler Lax pair system
\begin{align}
    d\momega&= \momega \circ \momega\,, &
    d\mC&= \momega\circ \mC-\mC \circ \momega\,,
\end{align}
where the fields take values in the deformed algebra. There exists an explicit formula $\omega[\momega,\mC]$, $C[\mC]$ for the solutions of \eqref{problemgen} \cite{Sharapov:2019vyd}. We note that the Lax pair equations are well-defined and do not have any locality problem. The explicit formula for the solutions of the original system \eqref{problemgen} allows one to avoid dealing with nonlocalities whatsoever. These facts give an importance to the deformed higher spin algebra as the only structure controlling interactions in formal higher spin gravities. 

Since the computation along the canonical AdS/CFT path is expected to face the problem of nonlocalities in higher spin theories, it is important to come up with a set of well-defined observables. As was suggested in \cite{Sezgin:2011hq} a natural set of observables is given by invariants of higher spin symmetries. The simplest invariants are traces of the master field $\mC$   
\begin{align}
    \mathrm{Tr}(\mC \circ...\circ \mC)\,.
\end{align}
It is important that when the deformation is switched off the invariants reduce to those of the initial higher spin algebra and are known to reproduce the correct correlation functions
\cite{Colombo:2012jx,Didenko:2013bj,Didenko:2012tv,Bonezzi:2017vha}.
A larger set of invariants is discussed in
\cite{Sharapov:2019vyd}.

\section{\label{sec:deformation}Deformation}
The problem of constructing the interaction vertices therefore reduces to that of deforming the smash-product algebra $\hsdouble$. This is a much simpler problem to solve\footnote{For example, the case of the smashed product of the Weyl algebra with any group of symplectic reflections was studied in \cite{Sharapov:2017lxr,Sharapov:2018hnl}. } and one can also present certain model-independent arguments that such deformation is always possible \cite{Sharapov:2018kjz}. Nevertheless, the $5d$ case resisted a number of attempts over the years. It turns out that the usual oscillator realization \eqref{usualosc} does not admit the deformation we are looking for. One can easily realize the $\pi$-automorphism as the reflection of four (out of eight) oscillators $a_A$, $b^B$, but the corresponding smash-product algebra $A_4\rtimes \mathbb{Z}_2$ does not have any nontrivial deformations \cite{AFLS,Sharapov:2017lxr,Sharapov:2018hnl} since its second Hochschild cohomology group vanishes.\footnote{This is not in contradiction with the fact that $\hsdouble$ should admit the deformation. $\hsdouble$ is realized as a subquotient \eqref{subquotient} of the Weyl algebra $A_4$ ($4$ stands for the number of canonical pairs, $a_A$, $b^B$). Therefore, there is no direct relation between the rigidity of $A_4\rtimes \mathbb{Z}_2$ and $\hsdouble$ being soft. This explains why the methods based on \eqref{subquotient} are not applicable here. } Therefore, we have to resort to other tools.

The problem now is to describe the deformation of $\hsdouble$. We found two ways to achieve that. The first one, section \ref{sec:defjoseph}, is to directly deform the Joseph relations and hence the the quotient of $U(su(2,2)) \rtimes\mathbb{Z}_2$. The second one, section \ref{sec:quasiconformal}, is to deform the quasi-conformal realization of the higher spin algebra.

\subsection{\label{sec:defjoseph}Deformation Through the Universal Enveloping Algebra}
It is convenient to describe $\hsdouble$ in the Lorentz base where the $su(2,2)$ commutation relations have the form \eqref{LLLPPP}. To obtain $\hsdouble$ we add one more generator  $k$, $k^2=1$, such that $kLk=L$ and $kPk=-P$. The set of Joseph's relations \eqref{Joseph} split into the triple of finite-dimensional irreducible modules of $su(2,2)$: the trivial module corresponding to the Casimir operator, the $15$-dimensional adjoint representation $(1,0,1)$, and the $20$-dimensional representation  of weight $(0,2,0)$. The value of the Casimir operator is fixed by the self-consistency of the ideal.\footnote{One can obtain it by reducing the element  $T\fdu{A}{B}T\fdu{C}{D}T\fdu{E}{F}$ in two different ways \cite{Eastwood:2002su}.} Since the modules are irreducible one can take the following two Lorentz components of the nontrivial modules as the `lowest weight vectors' (we omit $\star$ below):
\begin{align}\label{JosephLorentz}
    I&\equiv\frac12 P_{AB}P^{AB}-m^2\,,&
    I_{AB}&\equiv \{L_{AM},P\fdu{B}{M}\}+\{L_{BM},P\fdu{A}{M}\}\,.
\end{align}
The rest of the relations are obtained by commuting these two with $P_{AB}$. Note that $15=10+5$ and $20=14+5+1$ as representations of $sp(4)$. The consistency fixes $m^2=-2$ (in the units of the cosmological constant that we set to $1$). Note that \eqref{LLLPPP} and \eqref{JosephLorentz} give a complete description of $\hs$ and $\hsdouble$. It is possible to work with the free equations directly in terms of fields taking values in the universal enveloping algebra $U(so(d,2))$ modulo the Joseph relations \cite{Iazeolla:2008ix}.

First of all, the very definition of $\hs$ as of the universal enveloping algebra modulo the Joseph relations suggests that the sought for deformation of $\hsdouble$ can be described in the same language. The starting point is to keep \eqref{LL} and \eqref{LP}, but deform \eqref{PP} into\footnote{This type of a deformed commutator has already appeared in \cite{Sharapov:2018kjz,Sharapov:2019vyd} for different higher spin algebras. One of the main findings of the present paper is that this is the only relation that completely determines the nonlinear equations of motion.  }
\begin{align}\label{PPdeformed}
    [P^{AB},P^{CD}]&= (1+\nu k) (L^{AD}C^{BC}-L^{BD}C^{AC}-L^{AC}C^{BD}+L^{BC}C^{AD})\,.
\end{align}
It is this modification that drives the whole deformation. It also leads to the correct Einstein's equations as we show below. The requirement not to deform the $[L,L]$ and $[L,P]$ commutators is a form of the equivalence principle: we ought to preserve the local Lorentz algebra and its action on the tensors. Otherwise, the tensorial interpretation of the components of $\omega$ and $C$ is lost. We keep the `lowest weight vectors' \eqref{JosephLorentz} the same, but let $m^2$ depend on $k$. Indeed, there is no other deformation possible for \eqref{JosephLorentz}, which explains our choice. Acting with $[P_{AB},\bullet ]$, we generate the other components of the deformed Joseph's ideal:
\begin{align}\notag
    (0,0,0)&: &&
    \left\{\begin{aligned}
        &2C_2=\frac12P_{AB}P^{AB}-\frac12L_{AB}L^{AB}=- \frac12(6+ \nu k)(2+\nu k)\,;
    \end{aligned}\right.\\
    (0,2,0)&: && \label{deformedJoseph}
    \left\{\begin{aligned}
        &\frac12 P_{AB}P^{AB}-m^2=0\,,\\
        &\{L_{AM},P\fdu{B}{M}\}-\{L_{BM},P\fdu{A}{M}\}-2\nu k P_{AB}=0\,,\\
        &(\{L\fdu{[A}{[B}, L\fdu{C]}{D]}\}+\{P\fdu{[A}{[B}, P\fdu{C]}{D]}\})+\\
        &\qquad+2\nu k(2+\nu k) C_{AC}C^{BD}-(2+\nu k)^2(\delta\fdu{A}{B}\delta\fdu{C}{D}\!\!-\delta\fdu{C}{B}\delta\fdu{A}{D})=0\,;
    \end{aligned}\right.\\
    (1,0,1)&: &&\left\{
    \begin{aligned}
        &\{L_{AM},P\fdu{B}{M}\}+\{L_{BM},P\fdu{A}{M}\}=0\,,\\
        &\{L_{AM},L\fdu{B}{M}\}+\frac12 C_{AB}(2+\nu k)(-4+\nu k)=0\,.
    \end{aligned}\right. \notag
\end{align}
Self-consistency also requires $m^2=-(2+\nu k)(1+\nu k)$. 
Upon setting $\nu=0$, we recover the original Joseph's relations that are equivalent to \eqref{Joseph}. 
Even though $su(2,2)$ gets deformed by \eqref{PPdeformed}, there is still a sense in which we have the same $1+15+20$ as the total number of relations (the decomposition into Lorentz tensors still makes sense).

Much as the standard Joseph's relations determine the higher spin algebra $\hs$, the deformed Joseph's relations \eqref{deformedJoseph}, together with \eqref{LL}, \eqref{LP} and the deformed commutator \eqref{PPdeformed}, determine the deformation of $\hsdouble$. 
The deformation is smooth in the sense that for any two elements $f(L,P,k)$ and $g(L,P,k)$ we can, as a matter of principle, compute their product $f\circ g$ and decompose it into irreducible Lorentz tensors. Therefore, the deformed algebra is well-defined and the vertices can be written down.

It is easy to read off the Lorentz spectrum of the (un)deformed algebra directly from \eqref{deformedJoseph}. First of all, there are no singlets except for the unit element itself since $P^2$, $L^2$ are equivalent to ($k$-dependent) numbers. Secondly, all single contractions, $L_{AM}P\fdu{B}{M}$, $L_{AM}L\fdu{B}{M}$, and  $P_{AM}P\fdu{B}{M}$, can be transformed into $L_{AB}$, $P_{AB}$ and $C_{AB}$. Lastly, the four-index relation implies that the two ways of getting $(0,2)$ of $sp(4)$ via appropriate projections of $L_{AB}L_{CD}$ and $P_{AB}P_{CD}$ are equivalent. Therefore, the spectrum of the algebra consists of $sp(4)$-tensors of weight $(2k,m)$, $m,k=0,1,...$ that can be thought of as coefficients of the appropriately symmetrized monomials $L^kP^m$. The spectrum is multiplicity free. The Young diagram of $(2k,m)$ has $2k+m$ boxes in the first row and $m$ boxes in the second.

In practice, one may want to have an efficient tool to compute the $\circ$-product of any two elements.\footnote{It seems that the efficient methods to compute the structure constants of \cite{Iazeolla:2008ix,Joung:2014qya} can be extended from higher spin algebras to their deformations.} Any (e.g., oscillator) realization of the deformed algebra must fulfill the deformed Joseph's relations we derived above. 

\subsection{\label{sec:quasiconformal}Quasi-conformal Realization}
While the widely used oscillator realization \eqref{subquotient} is rigid, it is clear that the quasi-conformal realization, being  minimal, must admit a deformation that we constructed in the previous section. The automorphism $\pi$ is realized here in a very simple way: 
\begin{align}
\pi(p_z)=-p_z\,,& & \pi(z)=-z\,,
\end{align}
leaving all the other generators intact. The corresponding smash-product algebra is defined by the relations 
\begin{align}\label{kzp}
    k^2&=1\,, && \{z,k\}=0\,, && \{p_z,k\}=0\,,
\end{align}
along with the requirement that  $k$ commutes to $x,p_x,y,p_y$. The desired deformation can now obtained by  redefining  the `momentum' 
\begin{align}\label{pz}
    p_z\quad \mapsto \quad \tilde p_z=p_z+ \frac{i\nu }{2z}k\,.
\end{align}
Upon this redefinition the (anti)commutation relations for the triple $z, \tilde p_z, k$ take the form 
\begin{align}\label{defosc}
    [z,\tilde p_z]&=i(1+\nu k)\,, && \{z,k\}=0\,, && \{\tilde p_z,k\}=0\,.
\end{align}
These relations are known as the `deformed oscillator algebra' \cite{Yang:1951pyq}. The change of variables (\ref{pz}) should be accompanied with the following change of the composite operators $Y^A_L$:
\begin{align}\label{deformedY}
    Y_L^A&=\Big\{z, \tilde p_z, 0, 
    \frac{1}{x}( z p_y-\tilde p_z y-\tfrac12-\tfrac12 \nu k) \Big\}\,,
\end{align}
while all $Y^A_R$ stay the same. It is now a simple exercise in algebra to see that the Lorentz $L_{AB}$ and translations $P_{AB}$ generators, being defined by the same formulas \eqref{sufromY}, \eqref{LorTrans}, do fulfill the deformed Joseph's relations \eqref{deformedJoseph} together with \eqref{LL}, \eqref{LP}, and \eqref{PPdeformed}. This gives an explicit quasi-conformal realization of the deformed algebra $\hsdouble$. Thus, Rels. \eqref{defosc}, \eqref{deformedY} provide a complete solution of the $5d$ higher spin problem. It is interesting to note, that an obviously  nontrivial deformation of the extended higher spin algebra  generated by $L_{AB}$, $P_{AB}$, and $k$ is induced by the trivial deformation (\ref{pz}) of the algebra of rational functions in non-commuting variables $z,p_z,x,p_x,y,p_y$ and $k$.  

\section{\label{sec:Einstein}Einstein Equations}
Let us clarify the origin of the $[P,P]$-commutator \eqref{PPdeformed}, which plays the role of a seed that drives the whole deformation. It results in the Einstein equations and is, therefore, well-motivated. To show this, let us switch off all the higher spin components in $\omega$ and $C$ and concentrate on the spin-two sector.\footnote{It is well to bear in mind that this is not a consistent truncation of the full theory: since each field sources all the others, one cannot just set higher spin fields to zero.} The Einstein equations are realized as the $P_{AB}$ and $L_{AB}$ components of \eqref{omegaeq}:
\begin{align*}
    P_{AB}&: & dh^{AB}-\varpi\fud{[A}{C}\wedge h^{CB]}&=0\,,\\
    L_{AB}&: & d\varpi^{AB}-\varpi\fud{A}{C}\wedge \varpi^{CB}&=h\fud{A}{C}\wedge h^{CB}+ h\fdu{C}{M}\wedge h_{MD} W^{ABCD}\,,
\end{align*}
where $W$ has the symmetries of the Weyl tensor.\footnote{The Weyl tensor in the $so(4,1)$ language $W_{ab,cd}$ has the same symmetries as the Riemann tensor and is also traceless. In the $sp(4)$ language this corresponds to a rank-four totally symmetric tensor $W^{ABCD}$. } It is embedded into $C$ as $W^{ABCD}L_{AB}L_{CD}$. The first equation is the torsion constraint. The second one implies that the Riemann two-form (the left-hand side) consists of the cosmological $hh$-term (it comes from the undeformed $[P,P]$-commutator) and from the Weyl tensor. Thus, the traceless part of the Ricci tensor vanishes. This is a fancy yet correct way to impose the vacuum Einstein's equations. Let us see how the Weyl tensor emerges from the deformation. We write, omitting irrelevant terms and numerical factors, 
\begin{align*}
    &\mathcal{V}_3(\omega,\omega,C)= h_{MC} \wedge h_{ND}\,\phi_1(P^{MC},P^{ND})\star Ck\sim
    h\fdu{C}{M}\wedge h_{MD}( k L^{CD})\, (W^{ABEF}L_{AB}L_{EF})k\\[3mm]
    &\qquad\quad \sim h\fdu{C}{M}\wedge h_{MD}\,\delta\fdu{A}{C}\delta\fdu{B}{D}\, W^{ABEF}L_{EF}\sim h\fdu{C}{M}\wedge h_{MD} W^{ABCD} L_{AB}\,.
\end{align*}
Here, we used \eqref{cubic} with $C\rightarrow Ck$; $\phi_1$ results from \eqref{PPdeformed}. In the last line we used the Joseph relations that imply that $L_{AB}L^{AB}=8+\mathcal{O}(\nu)$ and hence the product $\{L_{AB},L_{CD}\}$ contains the singlet component in addition the others, that is, $\{L_{AB},L_{CD}\}=C_{AC}C_{BD}+\ldots$. It is this singlet that reduces the power of $L$'s from $3$ to $1$, generating the proper right hand side for the Riemann two-form. Note that the potentially dangerous $\varpi\varpi W$ and $\varpi h W$ terms vanish since the deformation preserves both $[L,L]$ and $[L,P]$ commutators.

In the $so(4,1)$ language the Lorentz and translation generators are $L_{ab}=-L_{ba}$ and $P_a$, $a,b=1,\ldots,5$. We can summarize the physical interpretation of the deformation \eqref{PPdeformed} as follows (now it is written in a $d$-independent form):
\begin{align}\label{PPdeformedA}
    [P_a,P_b]&= (1+\nu k )L_{ab} && \Longleftrightarrow && \text{(Einstein Equations)}\,.
\end{align}
Of course, the Einstein equations is a part of numerous non higher spin theories. The difference is that the deformation \eqref{PPdeformedA} is a small part of the Hochschild cocycle $\phi_1$ of the higher spin algebra and leads to the $A_\infty$-algebra eventually \cite{Sharapov:2018kjz,Sharapov:2019vyd}. It is the associative structure of the higher spin algebras and of the deformation they lead to that brings higher spin fields in. In the lower-spin theories containing gravity, the same commutator $[P_a,P_b]$ gets deformed, but only as the Chevalley--Eilenberg cocycle of the Poincar{\'e} or (anti)-de Sitter algebra.\footnote{It becomes more transparent when formulated in the frame-like language of vielbien and spin-connection, which can be interpreted as the gauge fields of the symmetry algebras of the maximally symmetric vacua.} As such, it does not call for  a higher spin extension.

\section{\label{sec:conclusions}Conclusions}
In the paper we constructed the bosonic formal higher spin gravity in $AdS_5$. The cornerstone of any formal Higher Spin Gravity is a higher spin algebra. The definition of higher spin algebras via enveloping algebras and Joseph's relations is a fundamental one. We suggested a new way to construct the interaction vertices by deforming the Joseph ideal relations as well as the commutator of the $AdS_5$ translations. This simplifies significantly the construction and puts the full nonlinear theory on the same algebraic ground as the higher spin algebra. 

Any other realization of the $5d$ theory has to fulfill the deformed Joseph's relations we found. The usual oscillator realization does not admit the deformation and cannot be used to construct interactions. Instead, we constructed a quasi-conformal realization of the deformed algebra. Obviously, the two approaches -- through the universal enveloping algebra and the quasi-conformal realization -- are applicable to all other higher spin gravities, both known and yet to be constructed. Some examples of interest include the $5d$ supersymmetric theories, the $7d$ higher spin gravities \cite{Sezgin:2001ij}, and the $6d$ exceptional higher spin gravity based on the $F(4)$ superalgebra \cite{Gunaydin:2016amv}.

The considered bosonic theory opens up the way for the supersymmetric extensions, where $su(2,2)$ gets replaced by $su(2,2|N)$. The massless sector of tensionless strings should be described by a theory based on the higher spin extension of $psu(2,2|4)$ \cite{Sezgin:2001yf}. We expect that the two approaches presented here should admit a straightforward supersymmetric extension, e.g. the quasi-conformal realization is available \cite{Fernando:2009fq,Govil:2013uta}. Contrary to the case of $AdS_4$ there is an upper bound $\mathcal{N}\leq 8$ on the number of supersymmetries for higher spin gravities in $AdS_5$.

It is worth mentioning, that there are not so many ways to deform enveloping algebras. Well-known is the quantum deformations of the Hopf structure. We seem to have found another way: enveloping algebras evaluated in certain irreducible representations do admit a deformation as associative algebras once extended with automorphisms (while the quotient algebras themselves may not have any deformations). This is closely related to the quantization of the corresponding Poisson orbifolds. 

Lastly, it would be interesting to establish a more precise relation between integrable quantum-mechanical models, e.g. the Calogero model, and formal Higher Spin Gravities.

\section*{Acknowledgments}
\label{sec:Aknowledgements}
We are grateful to Murat Gunaydin, Karapet Mkrtchyan, Ergin Sezgin and Per Sundell for very useful discussions. The work of E.S. was supported by the Russian Science Foundation grant 18-72-10123 in association with the Lebedev Physical Institute. The work of T.T. is supported by the International Max Planck Research School for Mathematical and Physical Aspects of Gravitation, Cosmology and Quantum Field Theory.

\footnotesize
\providecommand{\href}[2]{#2}\begingroup\raggedright\endgroup


\begin{thebibliography}{10}

\bibitem{Weinberg:1964ew}
S.~Weinberg, ``{Photons and Gravitons in S Matrix Theory: Derivation of Charge
  Conservation and Equality of Gravitational and Inertial Mass},''
\href{http://dx.doi.org/10.1103/PhysRev.135.B1049}{{\em Phys. Rev.} {\bfseries
  135} (1964) B1049--B1056}.

\bibitem{Coleman:1967ad}
S.~R. Coleman and J.~Mandula, ``{All Possible Symmetries of the S Matrix},''
\href{http://dx.doi.org/10.1103/PhysRev.159.1251}{{\em Phys. Rev.} {\bfseries
  159} (1967) 1251--1256}.

\bibitem{Bekaert:2010hp}
X.~Bekaert, N.~Boulanger, and S.~Leclercq, ``{Strong obstruction of the
  Berends-Burgers-van Dam spin-3 vertex},''
  \href{http://dx.doi.org/10.1088/1751-8113/43/18/185401}{{\em J.Phys.}
  {\bfseries A43} (2010) 185401},
\href{http://arxiv.org/abs/1002.0289}{{\ttfamily arXiv:1002.0289 [hep-th]}}.

\bibitem{Maldacena:2011jn}
J.~Maldacena and A.~Zhiboedov, ``{Constraining Conformal Field Theories with A
  Higher Spin Symmetry},''
\href{http://arxiv.org/abs/1112.1016}{{\ttfamily arXiv:1112.1016 [hep-th]}}.

\bibitem{Boulanger:2013zza}
N.~Boulanger, D.~Ponomarev, E.~Skvortsov, and M.~Taronna, ``{On the uniqueness
  of higher-spin symmetries in AdS and CFT},''
\href{http://arxiv.org/abs/1305.5180}{{\ttfamily arXiv:1305.5180 [hep-th]}}.

\bibitem{Alba:2013yda}
V.~Alba and K.~Diab, ``{Constraining conformal field theories with a higher
  spin symmetry in d=4},''
\href{http://arxiv.org/abs/1307.8092}{{\ttfamily arXiv:1307.8092 [hep-th]}}.

\bibitem{Alba:2015upa}
V.~Alba and K.~Diab, ``{Constraining conformal field theories with a higher
  spin symmetry in $d> 3$ dimensions},''
\href{http://arxiv.org/abs/1510.02535}{{\ttfamily arXiv:1510.02535 [hep-th]}}.

\bibitem{Bekaert:2015tva}
X.~Bekaert, J.~Erdmenger, D.~Ponomarev, and C.~Sleight, ``{Quartic AdS
  Interactions in Higher-Spin Gravity from Conformal Field Theory},''
  \href{http://dx.doi.org/10.1007/JHEP11(2015)149}{{\em JHEP} {\bfseries 11}
  (2015) 149},
\href{http://arxiv.org/abs/1508.04292}{{\ttfamily arXiv:1508.04292 [hep-th]}}.

\bibitem{Sleight:2017pcz}
C.~Sleight and M.~Taronna, ``{Higher-Spin Gauge Theories and Bulk Locality},''
  \href{http://dx.doi.org/10.1103/PhysRevLett.121.171604}{{\em Phys. Rev.
  Lett.} {\bfseries 121} no.~17, (2018) 171604},
\href{http://arxiv.org/abs/1704.07859}{{\ttfamily arXiv:1704.07859 [hep-th]}}.

\bibitem{Ponomarev:2017qab}
D.~Ponomarev, ``{A Note on (Non)-Locality in Holographic Higher Spin
  Theories},'' \href{http://dx.doi.org/10.3390/universe4010002}{{\em Universe}
  {\bfseries 4} no.~1, (2018) 2},
\href{http://arxiv.org/abs/1710.00403}{{\ttfamily arXiv:1710.00403 [hep-th]}}.

\bibitem{Blencowe:1988gj}
M.~Blencowe, ``{A Consistent Interacting Massless Higher Spin Field Theory in
  $D$ = (2+1)},''
\href{http://dx.doi.org/10.1088/0264-9381/6/4/005}{{\em Class.Quant.Grav.}
  {\bfseries 6} (1989) 443}.

\bibitem{Campoleoni:2010zq}
A.~Campoleoni, S.~Fredenhagen, S.~Pfenninger, and S.~Theisen, ``{Asymptotic
  symmetries of three-dimensional gravity coupled to higher-spin fields},''
  \href{http://dx.doi.org/10.1007/JHEP11(2010)007}{{\em JHEP} {\bfseries 1011}
  (2010) 007},
\href{http://arxiv.org/abs/1008.4744}{{\ttfamily arXiv:1008.4744 [hep-th]}}.

\bibitem{Henneaux:2010xg}
M.~Henneaux and S.-J. Rey, ``{Nonlinear $W_{infinity}$ as Asymptotic Symmetry
  of Three-Dimensional Higher Spin Anti-de Sitter Gravity},''
  \href{http://dx.doi.org/10.1007/JHEP12(2010)007}{{\em JHEP} {\bfseries 1012}
  (2010) 007},
\href{http://arxiv.org/abs/1008.4579}{{\ttfamily arXiv:1008.4579 [hep-th]}}.

\bibitem{Gaberdiel:2012uj}
M.~R. Gaberdiel and R.~Gopakumar, ``{Minimal Model Holography},''
  \href{http://dx.doi.org/10.1088/1751-8113/46/21/214002}{{\em J. Phys.}
  {\bfseries A46} (2013) 214002},
\href{http://arxiv.org/abs/1207.6697}{{\ttfamily arXiv:1207.6697 [hep-th]}}.

\bibitem{Segal:2002gd}
A.~Y. Segal, ``{Conformal higher spin theory},''
  \href{http://dx.doi.org/10.1016/S0550-3213(03)00368-7}{{\em Nucl. Phys.}
  {\bfseries B664} (2003) 59--130},
\href{http://arxiv.org/abs/hep-th/0207212}{{\ttfamily arXiv:hep-th/0207212
  [hep-th]}}.

\bibitem{Tseytlin:2002gz}
A.~A. Tseytlin, ``{On limits of superstring in AdS(5) x S**5},''
  \href{http://dx.doi.org/10.1023/A:1020646014240}{{\em Theor. Math. Phys.}
  {\bfseries 133} (2002) 1376--1389},
  \href{http://arxiv.org/abs/hep-th/0201112}{{\ttfamily arXiv:hep-th/0201112
  [hep-th]}}.
[Teor. Mat. Fiz.133,69(2002)].

\bibitem{Bekaert:2010ky}
X.~Bekaert, E.~Joung, and J.~Mourad, ``{Effective action in a higher-spin
  background},'' \href{http://dx.doi.org/10.1007/JHEP02(2011)048}{{\em JHEP}
  {\bfseries 02} (2011) 048},
\href{http://arxiv.org/abs/1012.2103}{{\ttfamily arXiv:1012.2103 [hep-th]}}.

\bibitem{Metsaev:1991nb}
R.~R. Metsaev, ``{S matrix approach to massless higher spins theory. 2: The
  Case of internal symmetry},''
\href{http://dx.doi.org/10.1142/S0217732391002839}{{\em Mod. Phys. Lett.}
  {\bfseries A6} (1991) 2411--2421}.

\bibitem{Metsaev:1991mt}
R.~R. Metsaev, ``{Poincare invariant dynamics of massless higher spins: Fourth
  order analysis on mass shell},''
\href{http://dx.doi.org/10.1142/S0217732391000348}{{\em Mod. Phys. Lett.}
  {\bfseries A6} (1991) 359--367}.

\bibitem{Ponomarev:2016lrm}
D.~Ponomarev and E.~D. Skvortsov, ``{Light-Front Higher-Spin Theories in Flat
  Space},'' \href{http://dx.doi.org/10.1088/1751-8121/aa56e7}{{\em J. Phys.}
  {\bfseries A50} no.~9, (2017) 095401},
\href{http://arxiv.org/abs/1609.04655}{{\ttfamily arXiv:1609.04655 [hep-th]}}.

\bibitem{Ponomarev:2017nrr}
D.~Ponomarev, ``{Chiral Higher Spin Theories and Self-Duality},''
  \href{http://dx.doi.org/10.1007/JHEP12(2017)141}{{\em JHEP} {\bfseries 12}
  (2017) 141},
\href{http://arxiv.org/abs/1710.00270}{{\ttfamily arXiv:1710.00270 [hep-th]}}.

\bibitem{Skvortsov:2018jea}
E.~D. Skvortsov, T.~Tran, and M.~Tsulaia, ``{Quantum Chiral Higher Spin
  Gravity},'' \href{http://dx.doi.org/10.1103/PhysRevLett.121.031601}{{\em
  Phys. Rev. Lett.} {\bfseries 121} no.~3, (2018) 031601},
\href{http://arxiv.org/abs/1805.00048}{{\ttfamily arXiv:1805.00048 [hep-th]}}.

\bibitem{Skvortsov:2018uru}
E.~Skvortsov, ``{Light-Front Bootstrap for Chern-Simons Matter Theories},''
  \href{http://dx.doi.org/10.1007/JHEP06(2019)058}{{\em JHEP} {\bfseries 06}
  (2019) 058},
\href{http://arxiv.org/abs/1811.12333}{{\ttfamily arXiv:1811.12333 [hep-th]}}.

\bibitem{Sundborg:2000wp}
B.~Sundborg, ``{Stringy gravity, interacting tensionless strings and massless
  higher spins},'' \href{http://dx.doi.org/10.1016/S0920-5632(01)01545-6}{{\em
  Nucl. Phys. Proc. Suppl.} {\bfseries 102} (2001) 113--119},
\href{http://arxiv.org/abs/hep-th/0103247}{{\ttfamily arXiv:hep-th/0103247}}.

\bibitem{Sezgin:2002rt}
E.~Sezgin and P.~Sundell, ``{Massless higher spins and holography},''
  \href{http://dx.doi.org/10.1016/S0550-3213(02)00739-3}{{\em Nucl.Phys.}
  {\bfseries B644} (2002) 303--370},
\href{http://arxiv.org/abs/hep-th/0205131}{{\ttfamily arXiv:hep-th/0205131
  [hep-th]}}.

\bibitem{Klebanov:2002ja}
I.~R. Klebanov and A.~M. Polyakov, ``{AdS dual of the critical O(N) vector
  model},'' \href{http://dx.doi.org/10.1016/S0370-2693(02)02980-5}{{\em Phys.
  Lett.} {\bfseries B550} (2002) 213--219},
\href{http://arxiv.org/abs/hep-th/0210114}{{\ttfamily arXiv:hep-th/0210114}}.

\bibitem{Heemskerk:2009pn}
I.~Heemskerk, J.~Penedones, J.~Polchinski, and J.~Sully, ``{Holography from
  Conformal Field Theory},''
  \href{http://dx.doi.org/10.1088/1126-6708/2009/10/079}{{\em JHEP} {\bfseries
  10} (2009) 079},
\href{http://arxiv.org/abs/0907.0151}{{\ttfamily arXiv:0907.0151 [hep-th]}}.

\bibitem{HaggiMani:2000ru}
P.~Haggi-Mani and B.~Sundborg, ``{Free large N supersymmetric Yang-Mills theory
  as a string theory},''
  \href{http://dx.doi.org/10.1088/1126-6708/2000/04/031}{{\em JHEP} {\bfseries
  04} (2000) 031},
\href{http://arxiv.org/abs/hep-th/0002189}{{\ttfamily arXiv:hep-th/0002189
  [hep-th]}}.

\bibitem{Gross:1988ue}
D.~J. Gross, ``{High-Energy Symmetries of String Theory},''
\href{http://dx.doi.org/10.1103/PhysRevLett.60.1229}{{\em Phys. Rev. Lett.}
  {\bfseries 60} (1988) 1229}.

\bibitem{Tseytlin:2002ny}
A.~A. Tseytlin, ``{Semiclassical quantization of superstrings: AdS(5) x S**5
  and beyond},'' \href{http://dx.doi.org/10.1142/S0217751X03012382}{{\em Int.
  J. Mod. Phys.} {\bfseries A18} (2003) 981--1006},
\href{http://arxiv.org/abs/hep-th/0209116}{{\ttfamily arXiv:hep-th/0209116
  [hep-th]}}.

\bibitem{Bonelli:2003zu}
G.~Bonelli, ``{On the covariant quantization of tensionless bosonic strings in
  AdS space-time},''
  \href{http://dx.doi.org/10.1088/1126-6708/2003/11/028}{{\em JHEP} {\bfseries
  11} (2003) 028},
\href{http://arxiv.org/abs/hep-th/0309222}{{\ttfamily arXiv:hep-th/0309222
  [hep-th]}}.

\bibitem{Eberhardt:2018ouy}
L.~Eberhardt, M.~R. Gaberdiel, and R.~Gopakumar, ``{The Worldsheet Dual of the
  Symmetric Product CFT},''
  \href{http://dx.doi.org/10.1007/JHEP04(2019)103}{{\em JHEP} {\bfseries 04}
  (2019) 103},
\href{http://arxiv.org/abs/1812.01007}{{\ttfamily arXiv:1812.01007 [hep-th]}}.

\bibitem{Gromov:2019aku}
N.~Gromov and A.~Sever, ``{The Holographic Fishchain},''
\href{http://arxiv.org/abs/1903.10508}{{\ttfamily arXiv:1903.10508 [hep-th]}}.

\bibitem{Eastwood:2002su}
M.~G. Eastwood, ``{Higher symmetries of the Laplacian},''
  \href{http://dx.doi.org/10.4007/annals.2005.161.1645}{{\em Annals Math.}
  {\bfseries 161} (2005) 1645--1665},
\href{http://arxiv.org/abs/hep-th/0206233}{{\ttfamily arXiv:hep-th/0206233
  [hep-th]}}.

\bibitem{Colombo:2012jx}
N.~Colombo and P.~Sundell, ``{Higher Spin Gravity Amplitudes From Zero-form
  Charges},''
\href{http://arxiv.org/abs/1208.3880}{{\ttfamily arXiv:1208.3880 [hep-th]}}.

\bibitem{Didenko:2013bj}
V.~E. Didenko, J.~Mei, and E.~D. Skvortsov, ``{Exact higher-spin symmetry in
  CFT: free fermion correlators from Vasiliev Theory},''
  \href{http://dx.doi.org/10.1103/PhysRevD.88.046011}{{\em Phys. Rev.}
  {\bfseries D88} (2013) 046011},
\href{http://arxiv.org/abs/1301.4166}{{\ttfamily arXiv:1301.4166 [hep-th]}}.

\bibitem{Didenko:2012tv}
V.~Didenko and E.~Skvortsov, ``{Exact higher-spin symmetry in CFT: all
  correlators in unbroken Vasiliev theory},''
  \href{http://dx.doi.org/10.1007/JHEP04(2013)158}{{\em JHEP} {\bfseries 1304}
  (2013) 158},
\href{http://arxiv.org/abs/1210.7963}{{\ttfamily arXiv:1210.7963 [hep-th]}}.

\bibitem{Bonezzi:2017vha}
R.~Bonezzi, N.~Boulanger, D.~De~Filippi, and P.~Sundell, ``{Noncommutative
  Wilson lines in higher-spin theory and correlation functions of conserved
  currents for free conformal fields},''
  \href{http://dx.doi.org/10.1088/1751-8121/aa8efa}{{\em J. Phys.} {\bfseries
  A50} no.~47, (2017) 475401},
\href{http://arxiv.org/abs/1705.03928}{{\ttfamily arXiv:1705.03928 [hep-th]}}.

\bibitem{Gaberdiel:1997ia}
M.~R. Gaberdiel and B.~Zwiebach, ``{Tensor constructions of open string
  theories. 1: Foundations},''
  \href{http://dx.doi.org/10.1016/S0550-3213(97)00580-4}{{\em Nucl. Phys.}
  {\bfseries B505} (1997) 569--624},
\href{http://arxiv.org/abs/hep-th/9705038}{{\ttfamily arXiv:hep-th/9705038
  [hep-th]}}.

\bibitem{Erler:2013xta}
T.~Erler, S.~Konopka, and I.~Sachs, ``{Resolving Witten`s superstring field
  theory},'' \href{http://dx.doi.org/10.1007/JHEP04(2014)150}{{\em JHEP}
  {\bfseries 04} (2014) 150},
\href{http://arxiv.org/abs/1312.2948}{{\ttfamily arXiv:1312.2948 [hep-th]}}.

\bibitem{Sharapov:2018kjz}
A.~Sharapov and E.~Skvortsov, ``{$A_\infty$ Algebras from Slightly Broken
  Higher Spin Symmetries},''
\href{http://arxiv.org/abs/1809.10027}{{\ttfamily arXiv:1809.10027 [hep-th]}}.

\bibitem{Sharapov:2019vyd}
A.~Sharapov and E.~Skvortsov, ``{Formal Higher Spin Gravities},''
  \href{http://dx.doi.org/10.1016/j.nuclphysb.2019.02.011}{{\em Nucl. Phys.}
  {\bfseries B941} (2019) 838--860},
\href{http://arxiv.org/abs/1901.01426}{{\ttfamily arXiv:1901.01426 [hep-th]}}.

\bibitem{Vasiliev:1990en}
M.~A. Vasiliev, ``Consistent equation for interacting gauge fields of all spins
  in (3+1)-dimensions,''
{\em Phys. Lett.} {\bfseries B243} (1990) 378--382.

\bibitem{Prokushkin:1998bq}
S.~Prokushkin and M.~A. Vasiliev, ``{Higher spin gauge interactions for massive
  matter fields in 3-D AdS space-time},''
  \href{http://dx.doi.org/10.1016/S0550-3213(98)00839-6}{{\em Nucl.Phys.}
  {\bfseries B545} (1999) 385},
\href{http://arxiv.org/abs/hep-th/9806236}{{\ttfamily arXiv:hep-th/9806236
  [hep-th]}}.

\bibitem{Vasiliev:2003ev}
M.~A. Vasiliev, ``{Nonlinear equations for symmetric massless higher spin
  fields in (A)dS(d)},''
  \href{http://dx.doi.org/10.1016/S0370-2693(03)00872-4}{{\em Phys. Lett.}
  {\bfseries B567} (2003) 139--151},
\href{http://arxiv.org/abs/hep-th/0304049}{{\ttfamily arXiv:hep-th/0304049
  [hep-th]}}.

\bibitem{Sagnotti:2005ns}
A.~Sagnotti, E.~Sezgin, and P.~Sundell, ``On higher spins with a strong sp(2,r)
  condition,''
\href{http://arxiv.org/abs/hep-th/0501156}{{\ttfamily hep-th/0501156}}.

\bibitem{Bonezzi:2016ttk}
R.~Bonezzi, N.~Boulanger, E.~Sezgin, and P.~Sundell,
  ``{Frobenius--Chern--Simons gauge theory},''
  \href{http://dx.doi.org/10.1088/1751-8121/50/5/055401}{{\em J. Phys.}
  {\bfseries A50} no.~5, (2017) 055401},
\href{http://arxiv.org/abs/1607.00726}{{\ttfamily arXiv:1607.00726 [hep-th]}}.

\bibitem{Bekaert:2017bpy}
X.~Bekaert, M.~Grigoriev, and E.~D. Skvortsov, ``{Higher Spin Extension of
  Fefferman-Graham Construction},''
  \href{http://dx.doi.org/10.3390/universe4020017}{{\em Universe} {\bfseries 4}
  no.~2, (2018) 17},
\href{http://arxiv.org/abs/1710.11463}{{\ttfamily arXiv:1710.11463 [hep-th]}}.

\bibitem{Arias:2017bvi}
C.~Arias, R.~Bonezzi, and P.~Sundell, ``{Bosonic Higher Spin Gravity in any
  Dimension with Dynamical Two-Form},''
  \href{http://dx.doi.org/10.1007/JHEP03(2019)001}{{\em JHEP} {\bfseries 03}
  (2019) 001},
\href{http://arxiv.org/abs/1712.03135}{{\ttfamily arXiv:1712.03135 [hep-th]}}.

\bibitem{Grigoriev:2018wrx}
M.~Grigoriev and E.~D. Skvortsov, ``{Type-B Formal Higher Spin Gravity},''
  \href{http://dx.doi.org/10.1007/JHEP05(2018)138}{{\em JHEP} {\bfseries 05}
  (2018) 138},
\href{http://arxiv.org/abs/1804.03196}{{\ttfamily arXiv:1804.03196 [hep-th]}}.

\bibitem{Sezgin:2001yf}
E.~Sezgin and P.~Sundell, ``{Towards massless higher spin extension of D=5, N=8
  gauged supergravity},''
  \href{http://dx.doi.org/10.1088/1126-6708/2001/09/025}{{\em JHEP} {\bfseries
  09} (2001) 025},
\href{http://arxiv.org/abs/hep-th/0107186}{{\ttfamily arXiv:hep-th/0107186
  [hep-th]}}.

\bibitem{Sezgin:2012ag}
E.~Sezgin and P.~Sundell, ``{Supersymmetric Higher Spin Theories},''
\href{http://arxiv.org/abs/1208.6019}{{\ttfamily arXiv:1208.6019 [hep-th]}}.

\bibitem{Fradkin:1989yd}
E.~S. Fradkin and V.~{\relax Ya}. Linetsky, ``{Conformal superalgebras of
  higher spins},''
\href{http://dx.doi.org/10.1016/0003-4916(90)90252-J}{{\em Annals Phys.}
  {\bfseries 198} (1990) 252--292}.

\bibitem{Fernando:2009fq}
S.~Fernando and M.~G{\"u}naydin, ``{Minimal unitary representation of SU(2,2)
  and its deformations as massless conformal fields and their supersymmetric
  extensions},'' \href{http://dx.doi.org/10.1063/1.3447773}{{\em J.Math.Phys.}
  {\bfseries 51} (2010) 082301},
\href{http://arxiv.org/abs/0908.3624}{{\ttfamily arXiv:0908.3624 [hep-th]}}.

\bibitem{Govil:2013uta}
K.~Govil and M.~G{\"u}naydin, ``{Deformed Twistors and Higher Spin Conformal
  (Super-)Algebras in Four Dimensions},''
  \href{http://dx.doi.org/10.1007/JHEP07(2014)004}{{\em JHEP} {\bfseries 03}
  (2015) 026},
\href{http://arxiv.org/abs/1312.2907}{{\ttfamily arXiv:1312.2907 [hep-th]}}.

\bibitem{Metsaev:2002vr}
R.~R. Metsaev, ``{Massless arbitrary spin fields in AdS(5)},''
  \href{http://dx.doi.org/10.1016/S0370-2693(02)01344-8}{{\em Phys. Lett.}
  {\bfseries B531} (2002) 152--160},
\href{http://arxiv.org/abs/hep-th/0201226}{{\ttfamily arXiv:hep-th/0201226
  [hep-th]}}.

\bibitem{Metsaev:2004ee}
R.~R. Metsaev, ``{Mixed symmetry massive fields in AdS(5)},''
  \href{http://dx.doi.org/10.1088/0264-9381/22/13/016}{{\em Class. Quant.
  Grav.} {\bfseries 22} (2005) 2777--2796},
\href{http://arxiv.org/abs/hep-th/0412311}{{\ttfamily arXiv:hep-th/0412311
  [hep-th]}}.

\bibitem{Metsaev:2014sfa}
R.~R. Metsaev, ``{Mixed-symmetry fields in AdS(5), conformal fields, and
  AdS/CFT},'' \href{http://dx.doi.org/10.1007/JHEP01(2015)077}{{\em JHEP}
  {\bfseries 01} (2015) 077},
\href{http://arxiv.org/abs/1410.7314}{{\ttfamily arXiv:1410.7314 [hep-th]}}.

\bibitem{Vasiliev:2001wa}
M.~A. Vasiliev, ``Cubic interactions of bosonic higher spin gauge fields in
  ads(5),'' {\em Nucl. Phys.} {\bfseries B616} (2001) 106--162,
\href{http://arxiv.org/abs/hep-th/0106200}{{\ttfamily hep-th/0106200}}.

\bibitem{Alkalaev:2002rq}
K.~B. Alkalaev and M.~A. Vasiliev, ``N = 1 supersymmetric theory of higher spin
  gauge fields in ads(5) at the cubic level,'' {\em Nucl. Phys.} {\bfseries
  B655} (2003) 57--92,
\href{http://arxiv.org/abs/hep-th/0206068}{{\ttfamily hep-th/0206068}}.

\bibitem{Alkalaev:2010af}
K.~Alkalaev, ``{FV-type action for $AdS_5$ mixed-symmetry fields},''
  \href{http://dx.doi.org/10.1007/JHEP03(2011)031}{{\em JHEP} {\bfseries 1103}
  (2011) 031},
\href{http://arxiv.org/abs/1011.6109}{{\ttfamily arXiv:1011.6109 [hep-th]}}.

\bibitem{Sezgin:2001zs}
E.~Sezgin and P.~Sundell, ``{Doubletons and 5D higher spin gauge theory},''
  {\em JHEP} {\bfseries 09} (2001) 036,
\href{http://arxiv.org/abs/hep-th/0105001}{{\ttfamily arXiv:hep-th/0105001}}.

\bibitem{Gunaydin:2000xr}
M.~Gunaydin, K.~Koepsell, and H.~Nicolai, ``{Conformal and quasiconformal
  realizations of exceptional Lie groups},''
  \href{http://dx.doi.org/10.1007/PL00005574}{{\em Commun. Math. Phys.}
  {\bfseries 221} (2001) 57--76},
\href{http://arxiv.org/abs/hep-th/0008063}{{\ttfamily arXiv:hep-th/0008063
  [hep-th]}}.

\bibitem{Mikhailov:2002bp}
A.~Mikhailov, ``{Notes on higher spin symmetries},''
\href{http://arxiv.org/abs/hep-th/0201019}{{\ttfamily arXiv:hep-th/0201019
  [hep-th]}}.

\bibitem{Joseph:1974hr}
A.~Joseph, ``{Minimal realizations and spectrum generating algebras},''
\href{http://dx.doi.org/10.1007/BF01646204}{{\em Commun. Math. Phys.}
  {\bfseries 36} (1974) 325--338}.

\bibitem{Vasiliev:1988sa}
M.~A. Vasiliev, ``Consistent equations for interacting massless fields of all
  spins in the first order in curvatures,''
{\em Annals Phys.} {\bfseries 190} (1989) 59--106.

\bibitem{Sezgin:2002ru}
E.~Sezgin and P.~Sundell, ``{Analysis of higher spin field equations in
  four-dimensions},''
  \href{http://dx.doi.org/10.1088/1126-6708/2002/07/055}{{\em JHEP} {\bfseries
  0207} (2002) 055},
\href{http://arxiv.org/abs/hep-th/0205132}{{\ttfamily arXiv:hep-th/0205132
  [hep-th]}}.

\bibitem{Sharapov:2019qqq}
A.~Sharapov and E.~Skvortsov, ``{ Cup product on $A_\infty$-cohomology and
  deformations},'' \href{http://arxiv.org/abs/1901.07872}{{\ttfamily
  arXiv:1901.07872 [math]}}.

\bibitem{Sezgin:2011hq}
E.~Sezgin and P.~Sundell, ``{Geometry and Observables in Vasiliev's Higher Spin
  Gravity},'' \href{http://dx.doi.org/10.1007/JHEP07(2012)121}{{\em JHEP}
  {\bfseries 07} (2012) 121},
\href{http://arxiv.org/abs/1103.2360}{{\ttfamily arXiv:1103.2360 [hep-th]}}.

\bibitem{Sharapov:2017lxr}
A.~A. Sharapov and E.~D. Skvortsov, ``{Hochschild cohomology of the Weyl
  algebra and Vasiliev's equations},'' {\em Letters in Mathematical Physics}
  {\bfseries 107} no.~12, (Dec, 2017) 2415--2432,
  \href{http://arxiv.org/abs/1705.02958}{{\ttfamily arXiv:1705.02958
  [math-ph]}}.

\bibitem{Sharapov:2018hnl}
A.~A. Sharapov and E.~D. Skvortsov, ``A simple construction of associative
  deformations,'' {\em Letters in Mathematical Physics} (Jul, 2018) ,
  \href{http://arxiv.org/abs/1803.10957}{{\ttfamily arXiv:1803.10957
  [math-ph]}}.

\bibitem{AFLS}
J.~Alev, M.~Farinati, T.~Lambre, and A.~Solotar, ``Homologie des invariants
  d'une algèbre de weyl sous l'action d'un groupe fini,'' {\em Journal of
  Algebra} {\bfseries 232} no.~2, (2000) 564--577.

\bibitem{Iazeolla:2008ix}
C.~Iazeolla and P.~Sundell, ``{A Fiber Approach to Harmonic Analysis of
  Unfolded Higher- Spin Field Equations},''
  \href{http://dx.doi.org/10.1088/1126-6708/2008/10/022}{{\em JHEP} {\bfseries
  10} (2008) 022},
\href{http://arxiv.org/abs/0806.1942}{{\ttfamily arXiv:0806.1942 [hep-th]}}.

\bibitem{Joung:2014qya}
E.~Joung and K.~Mkrtchyan, ``{Notes on higher-spin algebras: minimal
  representations and structure constants},''
  \href{http://dx.doi.org/10.1007/JHEP05(2014)103}{{\em JHEP} {\bfseries 05}
  (2014) 103},
\href{http://arxiv.org/abs/1401.7977}{{\ttfamily arXiv:1401.7977 [hep-th]}}.

\bibitem{Yang:1951pyq}
L.~M. Yang, ``{A Note on the Quantum Rule of the Harmonic Oscillator},''
\href{http://dx.doi.org/10.1103/PhysRev.84.788}{{\em Phys. Rev.} {\bfseries 84}
  no.~4, (1951) 788}.

\bibitem{Sezgin:2001ij}
E.~Sezgin and P.~Sundell, ``{7-D bosonic higher spin theory: Symmetry algebra
  and linearized constraints},''
  \href{http://dx.doi.org/10.1016/S0550-3213(02)00299-7}{{\em Nucl. Phys.}
  {\bfseries B634} (2002) 120--140},
\href{http://arxiv.org/abs/hep-th/0112100}{{\ttfamily arXiv:hep-th/0112100
  [hep-th]}}.

\bibitem{Gunaydin:2016amv}
M.~G{\"u}naydin, E.~D. Skvortsov, and T.~Tran, ``{Exceptional $F(4)$
  higher-spin theory in AdS$_{6}$ at one-loop and other tests of duality},''
  \href{http://dx.doi.org/10.1007/JHEP11(2016)168}{{\em JHEP} {\bfseries 11}
  (2016) 168},
\href{http://arxiv.org/abs/1608.07582}{{\ttfamily arXiv:1608.07582 [hep-th]}}.

\end{thebibliography}
\end{document}